\documentclass{SciPost}
\usepackage{graphicx}
\usepackage{amsmath}  
\usepackage{amsfonts} 
\usepackage{bm}

\begin{document}

\begin{center}{\Large \textbf{
Exactly solvable quantum few-body systems associated with the symmetries 
of the three-dimensional and four-dimensional icosahedra
}}\end{center}

\begin{center} 
T. Scoquart\textsuperscript{1,2},
J. J. Seaward\textsuperscript{2*},
S. G. Jackson\textsuperscript{3},
M. Olshanii\textsuperscript{2}
\end{center}

\begin{center}
{\bf 1} D\'{e}partement de Physique, Ecole Normale Sup\'{e}rieure, 24, rue Lhomond, 75005 Paris, France
\\
{\bf 2} Department of Physics, University of Massachusetts Boston, Boston, MA 02125, USA
\\
{\bf 3} Department of Mathematics, University of Massachusetts Boston, Boston Massachusetts 02125, USA
\\
* Joseph.Seaward001@umb.edu
\end{center}

\begin{center}
\today
\end{center}


\section*{Abstract}
{\bf 
The purpose of this article is to demonstrate that non-crystallographic reflection groups can be used 
to build new solvable  quantum particle systems. We explicitly construct a   one-parametric family 
of solvable four-body systems on a line, related to the 
symmetry of a regular icosahedron: in two distinct limiting cases the system is constrained to a half-line. We repeat the program 
for a $600$-cell, a four-dimensional generalization of the regular three-dimensional icosahedron. 
}

\vspace{10pt}
\noindent\rule{\textwidth}{1pt}
\tableofcontents\thispagestyle{fancy}
\noindent\rule{\textwidth}{1pt}
\vspace{10pt}

\section{Introduction}
\label{sec:intro}
The language of Lie groups that is traditionally employed when constructing new integrable quantum few- and many-body systems (\cite{gaudin1971_386}; \cite{gaudin1983_book}, Ch.\ 5 therein; \cite{Opdam_Yang's})
inadvertedly prohibits non-crystallographic symmetries from being considered, 
since no associated Lie groups exist.  
Some additional consistency can be gained 
if the context is shifted away from Lie groups and towards discrete reflection groups, affine or finite, classic or exceptional, 
crystallographic or not. In this paper, we explicitly construct two quantum solvable four- and five-body 
systems based on the non-crystallographic groups $H_{3}$ and $H_{4}$ respectively. 

We build on the general results obtained 
in the course of work devoted to extending the realm of integrable systems to the cases covered by 
the exceptional reflection groups \cite{olshanii2015_105005} (the case of $\tilde{F}_{4}$ in particular), 
long thought to be irrelevant (see \cite{gaudin1983_book}, paragraph 5.2.3(c) therein): prior to \cite{olshanii2015_105005}, the scope of applicable refection groups 
has been limited to $A_{N-1}$ (respectively $\tilde{A}_{N-1}$) and $C_{N}$ (resp.\ $\tilde{C}_{N}$) 
\cite{girardeau1960_516,lieb1963_1605,mcguire1963_622,gaudin1971_386}. These groups correspond to $N$ atoms of the 
same mass, on a line (resp.\ ring) 
and on a half-line (resp.\ in a box) respectively. 

The essence of the extension presented in \cite{olshanii2015_105005} is a diversification of the variety of maps 
between the particle coordinates and the Cartesian spaces in which the reflection groups operate. Such improvement 
allowed one to include ensembles of particles of different mass in consideration. As a result, it was possible to devise a 
general scheme according to which every reflection group---finite or affine---whose 
Coxeter diagram \cite{coxeter_book_regular_polytopes} does not have forks, generates an exactly solvable 
quantum few- or many-body system (or a few-parametric family of them) 
of hard-core particles on a line, a half-line, or in a box or on a ring.
We should note that when the associated reflection group is known, construction of the particle eigenstates 
\emph{per se} follows a known scheme that exists for any solvable kaleidoscopic cavity with homogeneous Robin
boundary conditions, irrespectively of 
whether it has a particle analogue or not \cite{gaudin1971_386,gutkin1979_6057,sutherland1980_1770,gutkin1982_1,gaudin1983_book,emsiz2006_191,emsiz2009_571,emsiz2010_61}.  

Regretfully, the above scheme does not allow for any extension to the case of finite strength interactions, if one 
requires the interactions be both of a two-body nature and act only on a contact. The physical 
reason is that the for finite interactions, particles are allowed to explore the whole multidimensional coordinate space where the reflection group operates. However, with the exception of the group $A_{N-1}$ (and $C_{N}$ with 
restrictions), the number 
of mirrors in the group grows much faster then the number of particle pairs. In Section.~4, we treat 
this phenomenon in more detail.

\section{$H_{3}$: symmetries of an icosahedron \label{sec:H3}}
Consider four hard-core particles on a line, with masses $m_{1}$, $m_{2}$, $m_{3}$, $m_{4}$, 
and coordinates $x_{1}$, $x_{2}$, $x_{3}$, $x_{4}$ respectively, with $x_{1}<x_{2}<x_{3}<x_{4}$. A coordinate transformation 
$x_{i} = \sqrt{\mu/m_{i}} z_{i}$ reduces the system to a four-dimensional particle of mass $\mu$. The
arbitrary mass scale $\mu$ is distinct from the total mass and can be chosen at will.  
The particle will be moving inside a hard-walled wedge formed by three hyperplanes of 
particle-particle contact with the outer 
normals 
\begin{align}
\begin{split}
&
\bm{\alpha}_{i}=\sqrt{m_{i}/(m_{i-1}+m_{i})} \bm{e}_{i-1} - \sqrt{m_{i-1}/(m_{i-1}+m_{i})} \bm{e}_{i}
\\
&
\mbox{for } i=2,\,3,\,4 \,\,,
\end{split}
\label{H3_simple_roots}
\end{align}
with $\bm{e}_{i}$ being unit vectors along the $z_{i}$-axes. The mutual 
orientation of the planes is not generic: these three planes cross along a line oriented along a 
unit vector $\bm{e}_{\mbox{\scriptsize COM}} \equiv \sum_{i=1}^{4} \sqrt{m_{i}/M} \bm{e}_{i}$, 
where $M \equiv  \sum_{i=1}^{4}  m_{i}$ is the total mass. Projection 
of the radius vector $\bm{z} \equiv (z_{1},\,z_{2},\,z_{3},\,z_{4})^{\top}$ onto the direction $\bm{e}_{\mbox{\scriptsize COM}}$ 
equals the position of the center of mass in the physical coordinates: 
$\bm{e}_{\mbox{\scriptsize COM}}\cdot \bm{z} = \sum_{i=1}^{4} m_{i} x_{i}/M$. For any set of masses, the time evolution of the center of mass coordinate $X_{\mbox{\scriptsize COM}} \equiv \bm{e}_{\mbox{\scriptsize COM}}\cdot \bm{z}$ can be separated from the rest of the dynamics.

\begin{figure}
\centering
\includegraphics[scale=.95]{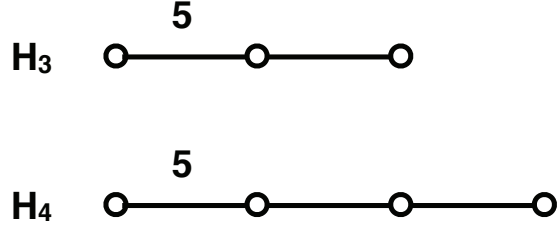}
\caption{
Coxeter diagrams \cite{coxeter_book_regular_polytopes} corresponding to the reflection groups $H_{3}$,  
i.e.\ the symmetry group of a regular icosahedron,
and $H_{4}$, i.e. the symmetry group of a $600$-cell, the four-dimensional 
cousin of a regular icosahedron.
The way the diagrams encode the relative orientation of the generating mirrors of the group 
is described both in the main text and in the caption to Fig.~\ref{f:sphere_COMPOUND_BB}.
             }
\label{f:coxeter_BB}        
\end{figure}

Dihedral angles between the plane of contact between $i-1$'st and $i$'th particles and its analogue for 
$i$'th and $i+1$'st particles  are given by \cite{mcguire1963_622}
%
\begin{align}
\theta_{(i-1)\,i\,(i+1)} = \arctan\sqrt{
                                                        \frac{
    m_{i}(m_{i-1}+m_{i}+m_{i+1})
                                                               }
                                                               {
   m_{i-1}m_{i+1}
                                                               }
                                              }
\,\,.
\label{tan_form}
\end{align}
For two non-overlapping pairs,  $m_{i-1}\mbox{-}m_{i}$ and $m_{j}\mbox{-}m_{j+1}$ with $j>i$, the corresponding 
hyperplanes are orthogonal to each other. 
Consider a full set of the particle-particle mirrors (three, for four particles). 
Some mirror arrangements form kaleidoscopes: in this case, the transformations of 
space caused by chains of sequential reflections form a finite 
group\footnote{In the case of particles on a ring or in a hard-wall box, the group is countably infinite.}. A complete 
list of these instances exists \cite{humphreys_book_1990,humphreys_book_1997}, and it is proven to be complete. Each instance of a 
kaleidoscopic mirror arrangement is encoded by a Coxeter diagram \cite{coxeter_book_regular_polytopes}. 
Fig.~\ref{f:coxeter_BB} provides examples of  Coxeter diagrams for the reflection groups $H_{3}$ and $H_{4}$.
Vertices correspond to the generating mirrors. Two vertices not connected by an edge correspond to two 
mirrors at a right angle between them. Two vertices connected by an unmarked edge give two mirrors at $60^{\circ}$
between them. Finally, edges labeled with an index $n$ produce a pair of mirrors at $(180/n)^{\circ}$. 

According to the
rules presented immediately above, the $H_{3}$ diagram at Fig.~\ref{f:coxeter_BB} 
produces three mirrors at angles $36^{\circ}$, $60^{\circ}$, and $90^{\circ}$
between them. It is easy to verify that each member of the following two-parametric family of the mass spectra 
produces such a set of particle-particle hyperplanes:
\begin{align}
\begin{split}
\left\{
\begin{array}{l}
m_{1} = \frac{ \xi + 1}{(5-2\sqrt{5})\xi -1} \, m_{2}
\\
m_{3} = \xi \,m_{2}
\\
m_{4} = \frac{\xi(\xi+1)}{3-\xi} \, m_{2}\,\,,
\end{array}
\right.
\begin{array}{l}
\mbox{with }
\frac{1}{5-2\sqrt{5}} \le \xi \le 3\,\,,
\\
\mbox{and } x_{1}<x_{2}<x_{3}<x_{4}
\,\,.
\end{array}
\end{split}
\label{H3_family}
\end{align}
The family is parametrized by an overall mass scale $m_{2} \ge 0$ and a dimensionless parameter $\xi$. The reason for 
the bounds on $\xi$ is the additional requirement of non-negativity of the masses involved. Two limiting cases deserve 
special attention,
$\xi \to (5-2\sqrt{5})^{-1}+0$ and $\xi \to 3-0$. In the first limit, the leftmost mass $m_{1}$ diverges. In a frame 
with the origin coinciding with the mass $m_{1}$ and co-moving 
with with it,  the problem reduces to a one-parametric family of three-body problems on a right half-line:
\begin{align*}
\begin{split}
\left\{
\begin{array}{l}
m_{3} = \frac{1}{5-2\sqrt{5}} \, m_{2}
\\
m_{4} = \left(\frac{5}{2} + \frac{11}{2\sqrt{5}}\right) \, m_{2}\,\,,
\end{array}
\right.
\begin{array}{l}
\mbox{with } m_{1}\to +\infty\mbox{, }x_{1}=0,
\\
\mbox{and } 0<x_{2}<x_{3}<x_{4}
\,\,.
\end{array}
\end{split}
\end{align*}
In the second limit, the rightmost mass $m_{4}$ diverges. Here, we obtain a one-parametric family of problems 
on a left half-line:
\begin{align*}
\begin{split}
\left\{
\begin{array}{l}
m_{1} = \frac{2}{7-3\sqrt{5}} \, m_2
\\
m_{3} = 3 \, m_{2}\,\,,
\end{array}
\right.
\begin{array}{l}
\mbox{with } m_{4}\to +\infty\mbox{, }x_{4}=0,
\\
\mbox{and } x_{1}<x_{2}<x_{3}<0
\,\,.
\end{array}
\end{split}
\end{align*}

\begin{figure}
    \centering
    (a)\!\!\!\!\!\!\!\!\!\!\!\!\!
        \includegraphics[scale=.65]{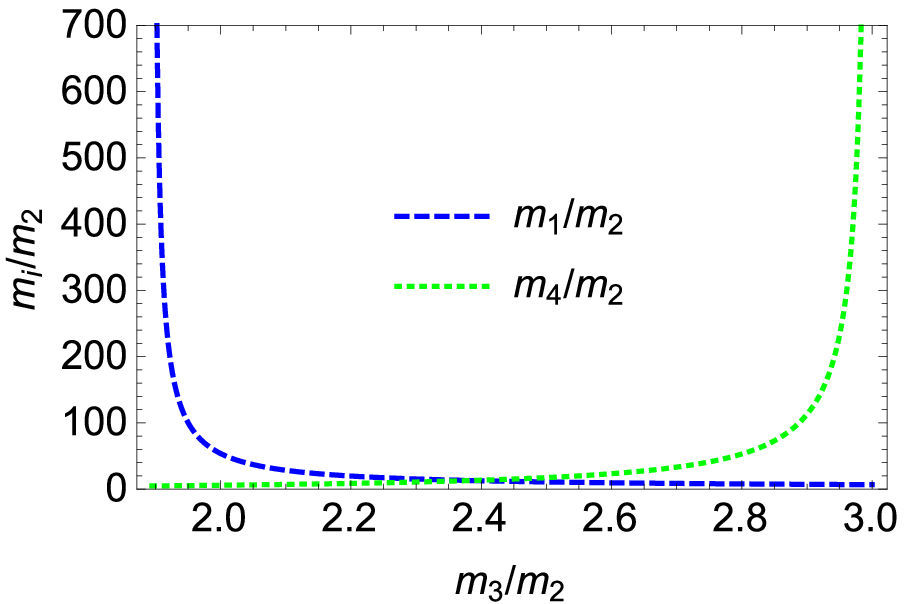}
        \qquad\qquad
    (b)\!\!\!\!\!\!\!\!\!\!\!\!\!
        \includegraphics[scale=.65]{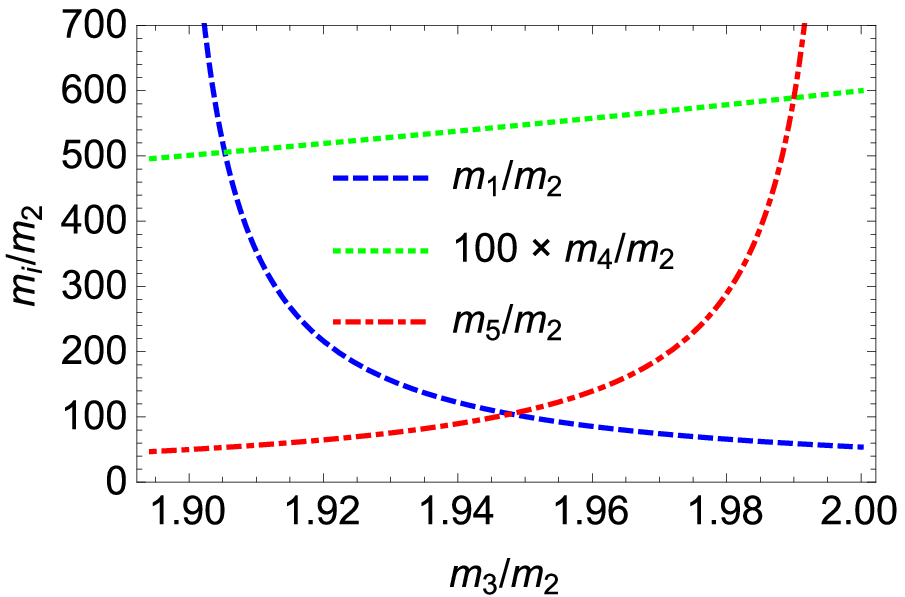}
    \caption{
    Mass ratios leading to icosahedral symmetries. $m_{i}$ is the mass of the $i$th particle in the sequence. 
    (a) A family of four-body mass spectra  
    that realize the reflection group $H_{3}$. (b)
    A family of five-body mass spectra  
    that realize the reflection group $H_{4}$. 
                 }
    \label{f:H3_and_H4_families}
\end{figure}

Fig.~\ref{f:H3_and_H4_families}(a)
illustrates the dependencies (\ref{H3_family}). 

When the hyperplanes of the particle-particle contact form a kaleidoscopic cavity, construction of the eigenstates 
becomes an easy task. In the case of $H_{3}$---the full symmetry group of an icosahedron---the sequential applications of reflections about the three 
generating mirrors (\ref{H3_simple_roots}) produce $120$ orthogonal transformations $\hat{g}$ that form 
this group. Eigenstates of the Hamiltonian, all of which are scattering states,  are parametrized by an incident 
wavevector $\bm{k}$:
\begin{align*}
&
\bm{\alpha}_{i} \cdot \bm{k} > 0
\mbox{ ,  for } i=2,\,3,\,4 \,\,:
\end{align*}
the corresponding Bethe Ansatz eigenstates
\cite{gaudin1971_386,gutkin1979_6057,sutherland1980_1770,gutkin1982_1,gaudin1983_book,emsiz2006_191,emsiz2009_571,emsiz2010_61} have a form
\begin{align}
\psi_{\bm{k}}(\bm{z}) = \mbox{const} \times \sum_{\hat{g}} (-1)^{{\cal P}(\hat{g})}  \exp[i (\hat{g}\bm{k})\cdot\bm{z}]
\,\,.
\label{H3_psi}
\end{align}
Here, $\mathcal{P}(\hat{g})$ is the parity of the group element $\hat{g}$: the parity of the number 
reflections about the generating mirrors (\ref{H3_simple_roots}) that lead to this element. 

\begin{figure}
\centering
\includegraphics[scale=.75]{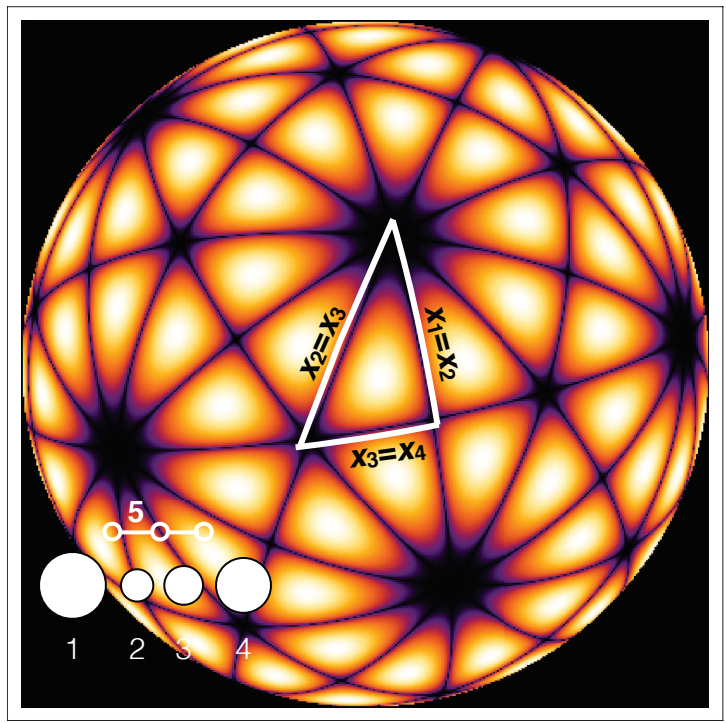}
\caption{
The white triangle bounds the physically allowed values of particle positions. In it, we show
the angular distribution, in the space of relative motion, 
of the probability density in the zero-energy state (\ref{H3_Q}) 
of four hard-core particles on a line, with a mass spectrum belonging to the family (\ref{H3_family}). 
Note that in this state, the angular distribution does not depend on the radial coordinate.  
A smooth continuation of this state to the 
remainder of the sphere is also shown, to illustrate the symmetry of the state. It is evident that the three angles 
of the ``physical'' triangle are  $36^{\circ}$, $60^{\circ}$, and $90^{\circ}$. 
These values are encoded in the Coxeter diagram (lower left 
corner) as the index $5$ (as in $36^{\circ}=\pi/5$) 
above the edge between the leftmost and the middle vertices,  ``empty'' index between 
the middle and  the rightmost vertices (the "$3$" in $\pi/3$ is omitted by convention,) and an ``empty'' edge between the leftmost and the rightmost vertices for the right angle (also omitted by convention.) Vertices 
themselves correspond to the sides of the triangle. From the particle perspective (labels and relative masses indicated below the Coxeter graph,) vertices of the 
Coxeter graph represent pairs of consecutive particles while the edges and the indices above them correspond the 
the consecutive particle triplets and the mass ratios in the triplet whose ratios are governed by (\ref{tan_form}). 
             }
\label{f:sphere_COMPOUND_BB}        
\end{figure}

For the problems with no bound states, scattering states of zero energy become the most fundamental object of interest. 
In the case of Bethe Ansatz states based on kaleidoscopic symmetries, the pure reflection members of the reflection
group---that also contains rotations and combinations of a rotation and a reflection---play the central role.
The group $H_{3}$ contains $15$ pure reflections, that correspond to $15$ symmetry planes of a regular 
icosahedron. Let $\bm{\beta}$ be one of the $15$ normals to the corresponding mirrors, where 
we assume, in order to avoid ambiguities, that $\bm{\alpha}_{i}\cdot\bm{\beta}>0$, for all $i=1,\,2,\,3$ and 
all $15$ normals $\bm{\beta}$ \footnote{Remark that according to an established terminology, the vectors 
opposite to $\bm{\alpha}_{i}$ and $\bm{\beta}_{j}$, i.e.\  $-\bm{\alpha}_{i}$ and $-\bm{\beta}_{j}$,
are the \emph{simple roots} and the \emph{positive roots} respectively.  
}. The normals $\bm{\beta}$ can be obtained by sequential applications of reflections about the generating mirrors to
a normal $\bm{\alpha}$ to one of them. It can be easily shown that 
the lowest degree anti-invariant polynomials of the corresponding group \cite{humphreys_book_1990},
\begin{align}
\psi_{\bm{k}=\bm{0}}({\bm z}) = \mbox{const} \times \prod_{\beta}  (\bm{\beta}\cdot{\bm z})
\,\,,
\label{H3_Q}
\end{align}
produce the desired zero-energy eigenstates of the problem\footnote{Since the Laplacian commutes with each element of the group, 
it would take the lowest degree homogeneous anti-invariant polynomial (\ref{H3_Q}) to a homogeneous 
anti-invariant polynomial of two degrees lower. However, by construction, there is no such polynomial \cite{humphreys_book_1990}. Thus, the action of the Laplacian must produce zero, i.e.\ (\ref{H3_Q}) must be a zero energy eigenstate of it.}.
Fig.~\ref{f:sphere_COMPOUND_BB} illustrates the probability density in the 
state (\ref{H3_Q}). Here, the position of the center of mass, 
$X_{\mbox{\scriptsize COM}} \equiv \sqrt{\mu/M} (\bm{e}_{\mbox{\scriptsize COM}}\cdot \bm{z})$,
is set to zero. In the residual three-dimensional subspace of the space the $\bm{z}$ coordinates belong to, 
the state (\ref{H3_Q}) factorizes into a product of a function of the radial coordinate 
$r = |\bm{z} - (\bm{e}_{\mbox{\scriptsize COM}}\cdot \bm{z})\bm{e}_{\mbox{\scriptsize COM}}|$ (that is proportional 
to $r^{15}$ in the $H_{3}$ case) and a function of angular coordinates. It is the latter  that is shown at 
Fig.~\ref{f:sphere_COMPOUND_BB}. 
%

\section{$H_{4}$: symmetries of a $600$-cell \label{sec:H4}}
$H_{4}$ is the full symmetry group of a $600$-cell \cite{coxeter_book_regular_polytopes}, a regular four-dimensional polytope (a four-dimensional Platonic solid)
that constitutes a four-dimensional analogue of a regular three-dimensional icosahedron. It three-dimensional 
``surface'' consists of regular tetrahedra, five meeting at each edge. 

In the case of the $H_{4}$ reflection group, one more mirror, at $60^{\circ}$ to the mirror corresponding 
to the rightmost vertex of the $H_{3}$ diagram (Fig.~\ref{f:coxeter_BB}) is added. Accordingly, 
a fifth particle is added to the system. The resulting two-parametric family of mass spectra is 
\begin{align}
\begin{split}
\left\{
\begin{array}{l}
m_{1} = \frac{ \xi + 1}{(5-2\sqrt{5})\xi -1} \, m_{2}
\\
m_{3} = \xi \,m_{2}
\\
m_{4} = \frac{\xi(\xi+1)}{3-\xi} \, m_{2}
\\
m_{5} = \frac{\xi(\xi+1)}{(3-\xi)(2-\xi)} \, m_{2} \,\,,
\end{array}
\right.
\begin{array}{l}
\mbox{with }
\frac{1}{5-2\sqrt{5}} \le \xi \le 2\,\,,
\\
\mbox{and } x_{1}<x_{2}<x_{3}<x_{4}<x_{5}
\,\,,
\end{array}
\end{split}
\label{H4_family}
\end{align}
where the two governing parameters are again a mass scale $m_{2}$ and a dimensionless ratio $\xi \equiv m_{3}/m_{2}$.
This family is illustrated at Fig.~\ref{f:H3_and_H4_families}(b).

Here, like in the $H_{3}$ case, we have two nontrivial special points. 
At $\xi \to (5-2\sqrt{5})^{-1}+0$, the mass spectrum converges to 
\begin{align*}
\begin{split}
\left\{
\begin{array}{l}
m_{3} = \frac{1}{5-2\sqrt{5}} \, m_{2}
\\
m_{4} = \left(\frac{5}{2} + \frac{11}{2\sqrt{5}}\right) \, m_{2}
\\
m_{5} = \left(\frac{47+21\sqrt{5}}{2}\right) \, m_{2}
\,\,,
\end{array}
\right.
\begin{array}{l}
\mbox{with } m_{1}\to +\infty\mbox{, }x_{1}=0,
\\
\mbox{and } 0<x_{2}<x_{3}<x_{4}<x_{5}
\,\,,
\end{array}
\end{split}
\end{align*}
and the system reduces to a four-body problem on a right half-line. Here, we are again assuming 
a moving frame whose origin coincides with the coordinate of the infinitely massive first particle at 
all times. 

The limit $\xi \to 2-0$ leads to a four-body problem on the left half-line:  
\begin{align*}
\begin{split}
\left\{
\begin{array}{l}
m_{1} = (27 + 12\sqrt{5})\, m_2
\\
m_{3} = 2 \, m_{2}
\\
m_{4} = 6 \, m_{2}
\,\,,
\end{array}
\right.
\begin{array}{l}
\mbox{with } m_{5}\to +\infty\mbox{, }x_{5}=0,
\\
\mbox{and } x_{1}<x_{2}<x_{3}<x_{4}<0
\,\,.
\end{array}
\end{split}
\end{align*}

For each member of the $H_{4}$ family of mass spectra (\ref{H4_family}), the outer 
normals to the generating mirrors are given by 
\begin{align}
\begin{split}
&
\bm{\alpha}_{i}=\sqrt{m_{i}/(m_{i-1}+m_{i})} \bm{e}_{i-1} - \sqrt{m_{i-1}/(m_{i-1}+m_{i})} \bm{e}_{i}
\\
&
\mbox{for } i=2,\,3,\,4,\,5 \,\,,
\end{split}
\label{H4_simple_roots}
\end{align}
Eigenstates of the $H_{4}$ problem will contain many more plane waves than in the $H_{3}$ case. 
All possible sequences of reflections about the generating mirrors for the full symmetry group of the 
$600$-cell, that contains $14400$ orthogonal transformations $\hat{g}$. Exactly like in the three-dimensional 
case, none of the transformations affects the dynamics of the center-of-mass coordinate 
$X_{\mbox{\scriptsize COM}} \equiv \sqrt{\mu/M} (\bm{e}_{\mbox{\scriptsize COM}}\cdot \bm{z})$, with 
$\bm{e}_{\mbox{\scriptsize COM}} \equiv \sum_{i=1}^{5} \sqrt{m_{i}/M} \bm{e}_{i}$ being 
the corresponding unit vector and $M \equiv  \sum_{i=1}^{5}  m_{i}$ being the total mass. There are 
$60$ pure reflections in the $H_{4}$, with the corresponding normals $\bm{\beta}$. 
The generic scattering states are given by the general formula (\ref{H3_psi}), with the sum 
running over all  $14400$ elements of the group, and 
$\bm{\alpha}_{i} \cdot \bm{k} > 0$, for $i=2,\,3,\,4,\,5$. The zero-energy scattering state will 
be again given by the expression (\ref{H3_Q}), where the product consists of $60$ factors,
and $\bm{\alpha}_{i}\cdot\bm{\beta}>0$, for all $i=1,\,2,\,3,\,4$ and 
all $60$ $\bm{\beta}$'s, to avoid ambiguity.


\section{Summary and outlook}
In this paper, we propose two new families of exactly solvable quantum four- and five-body problems;  these cases are 
associated with the symmetries of an icosahedron and a $600$-cell 
(i.e.\ a four-dimensional analogue of an icosahedron) respectively. 
This result explicitly demonstrates that non-crystallographic reflection groups can be used to construct quantum 
integrable few-body systems, on par with the crystallographic ones. In addition to the generic eigenstates we also analyze 
the zero-energy eigenstates, that correspond to the lowest-degree anti-invariant polynomials of the corresponding 
reflection group.    

We believe our results can not be extended to the case of finite-strength 
$\delta$-interactions between the particles if the local (i.e. contact) two-body 
nature of the interactions is to be preserved. Indeed it can be shown that, 
in order to preserve integrability of the system, the $\delta$-function potentials
at the 15 mirrors of the $H_{3}$ group must have the same strength, infinite or 
finite. In the finite case, any permutation of the four particles involved is possible, 
leading to 6 hyperplanes of contact, a number that differs from the number 
of mirrors in the group.
In contrast, for a given permutation, preserved over time, of the hard-core particles, 
only 3 hyperplanes of contact are physically accessible; the number of mirrors accessible,
given the particles' impenetrability, is also 3. Likewise, in the case of the $H_{4}$ group, the number
of the hyperplanes of particle-particle contact (i.e. 10) would differ from the number of the 
mirrors (i.e. 60). 

The relevance of the ``counting'' argument above can be strengthened by the most tangible 
case of two $\delta$-interacting particles in the field of a fixed $\delta$-potential of a different strength.
Here, in the two-dimensional plane of system's coordinate space, the potential is localized along the horizontal, vertical, and one of the 
diagonal lines, a set is clearly not closed under reflections  about its own members. And, as it is shown in \cite{morse1953_2_1709}, 
the eigenstates show features inconsistent with integrability, diffraction being the primary one. The mirror 
symmetry ($C_{2}$ in this case) and the associated integrability could be restored 
by adding an unphysical interaction that acts when the particles are located at the same distance from the potential but 
on the opposite sides of it. And finally the system can be returned to the realm of physical by raising the strength 
of the stationary potential to infinity, while keeping the "unphyisical" part of the interparticle interactions. For in an initial 
condition where both particles start at the same side of the potential, they would simply not be capable to explore the 
"unphyisical" part. In this example, both the empirical relevance and the integrability of a model can be preserved, but 
only at the expense of reducing choice of one of the interactions to infinite values. 

The remaining non-crystallographic reflection groups, $I_{2}(m)$, associated with the symmetries of 
regular polygons, 
deserve attention. Even though the resulting three body integrable systems---whose classical versions were 
analysed in \cite{hwang2015_467}---are conceptually much simpler then most of the other problems of this class, 
there are two aspects call for closer consideration. Firstly, as it has been shown classically in \cite{hwang2015_467}, 
a many-body system that contains integrable few-body sub-systems shows 
a slowdown of relaxation: a quantum
version of the phenomenon is in order. The case of $I_{2}(m)$ symmetry is the most empirically relevant, since it 
can be realized with \emph{only two} atomic species. Secondly, exact eigenstates, albeit not of the  
Bethe Ansatz type, can be obtained for any set of masses of three hard-core particles on a line; a separation 
of the radial and angular components of the relative motion can be used. This case can be used to analyze the relationship
between the Bethe Ansatz integrability (along with possible associated Liouville integrability \cite{olshanii2015_105005}) 
and the existence of the exact solutions in general.  

On a different front, the answer to the question of existence of particle realizations of reflection symmetries with the bifurcating 
Coxeter diagrams, $D_{n}$($\tilde{D}_{n}$), $\tilde{B}_{n}$, and $E_{6,\,7,\,8}$($\tilde{E}_{6,\,7,\,8}$),  
remains as elusive as ever.

\section*{Acknowledgements}
The authors thank Vanja Dunjko for help and comments. 


\paragraph{Funding information}
This work was supported by the US National Science Foundation Grant No.\ PHY-1402249, 
the Office of Naval Research Grant N00014-12-1-0400, and 
a grant from the {\it Institut Francilien de Recherche sur les Atomes Froids} (IFRAF). 
Financial support for TS provided by the Ecole Normale Sup\'{e}rieure
is also appreciated.



\begin{thebibliography}{10}
\providecommand{\url}[1]{\texttt{#1}}
\providecommand{\urlprefix}{URL }
\expandafter\ifx\csname urlstyle\endcsname\relax
  \providecommand{\doi}[1]{doi:\discretionary{}{}{}#1}\else
  \providecommand{\doi}{doi:\discretionary{}{}{}\begingroup
  \urlstyle{rm}\Url}\fi
\providecommand{\eprint}[2][]{\url{#2}}

\bibitem{gaudin1971_386}
M.~Gaudin,
\newblock \emph{Boundary energy of a bose gas in one dimension},
\newblock Physical Review \textbf{A24}, 386 (1971).

\bibitem{gaudin1983_book}
M.~Gaudin,
\newblock \emph{La fonction d'onde de Bethe},
\newblock Masson, Paris; New York (1983).

\bibitem{Opdam_Yang's}
G.~J. Heckman, and E.~M. Opdam,
\newblock \emph{Yang's system of particles and Hecke algebras}, 
\newblock Ann. Math. \textbf{145}(1), 139 (1997).

\bibitem{olshanii2015_105005}
M.~Olshanii and S.~G. Jackson,
\newblock \emph{An exactly solvable quantum four-body problem associated with
  the symmetries of an octacube},
\newblock New J. Phys. \textbf{17}, 105005 (2015).

\bibitem{girardeau1960_516}
M.~Girardeau,
\newblock \emph{Realationship between systems of impenetrable bosons and
  fermions in one dimension},
\newblock J. Math. Phys. \textbf{1}(6), 516 (1960).

\bibitem{lieb1963_1605}
E.~H. Lieb and W.~Liniger,
\newblock \emph{Exact analysis of an interacting {B}ose gas. {I}. {T}he general
  solution and the ground state},
\newblock Phys.\ Rev. \textbf{130}, 1605 (1963).

\bibitem{mcguire1963_622}
J.~B. McGuire,
\newblock \emph{Study of exactly soluble one-dimensional $n$-body problems},
\newblock J. Math. Phys. \textbf{5}, 622 (1963).

\bibitem{coxeter_book_regular_polytopes}
H.~S.~M. Coxeter,
\newblock \emph{Regular Polytopes},
\newblock Methuen \& CO. LTD., London (1948).

\bibitem{gutkin1979_6057}
E.~Gutkin and B.~Sutherland,
\newblock \emph{Completely integrable systems and groups generated by
  reflections},
\newblock Proc. Natl. Acad. Sci. USA \textbf{76}, 6057 (1979).

\bibitem{sutherland1980_1770}
B.~Sutherland,
\newblock \emph{Nondiffractive scattering: Scattering from kaleidoscopes},
\newblock J. Math. Phys. \textbf{21}, 1770 (1980).

\bibitem{gutkin1982_1}
E.~Gutkin,
\newblock \emph{Integrable systems with delta-potential},
\newblock Duke Math. J. \textbf{49}, 1 (1982).

\bibitem{emsiz2006_191}
E.~Emsiz, E.~M. Opdam and J.~V. Stokman,
\newblock \emph{Periodic integrable systems with delta-potentials},
\newblock Comm. Math. Phys. \textbf{261}, 191 (2006).

\bibitem{emsiz2009_571}
E.~Emsiz, E.~M. Opdam and J.~V. Stokman,
\newblock \emph{Trigonometric {C}herednik algebra at critical level and quantum
  many-body problems},
\newblock Sel. math., New ser. \textbf{14}, 571 (2009).

\bibitem{emsiz2010_61}
E.~Emsiz,
\newblock \emph{Completeness of the {B}ethe ansatz on {W}eyl alcoves},
\newblock Lett. Math. Phys. \textbf{91}, 61 (2010).

\bibitem{humphreys_book_1990}
J.~Humphreys,
\newblock \emph{Reflection groups and Coxeter groups},
\newblock Cambridge Unversity Press, Cambridge (1990).

\bibitem{humphreys_book_1997}
J.~Humphreys,
\newblock \emph{Introduction to Lie Algebras and Representation Theory},
\newblock Springer, New York (1997).

\bibitem{hwang2015_467}
Z.~Hwang, F.~Cao and M.~Olshanii,
\newblock \emph{Traces of integrability in relaxation of one-dimensional
  two-mass mixtures},
\newblock J. Stat. Phys. \textbf{161}, 467 (2015).

\bibitem{morse1953_2_1709}
P.~M~.Morse and H.~Feshbach,
\newblock \emph{Methods of Theoretical Physics},
vol.~2, Ch.~12, paragraph~12.3, 
\newblock McGraw-Hill, NY (1953).

\end{thebibliography}

\nolinenumbers

\end{document}